\documentclass[11pt]{article}
\usepackage{moriond,epsfig}

\def\NP{{\em Nucl. Phys.} }
\def\PLB{{\em Phys. Lett.}  B}
\def\PRL{\em Phys. Rev. Lett.}
\def\PRD{{\em Phys. Rev.} D}
\def\PR{{\em Phys. Rev.} }

\def\be{\begin{equation}}
\def\ee{\end{equation}}
\def\bea{\begin{eqnarray}}
\def\eea{\end{eqnarray}}

\begin{document}
%
  \begin{center}
    To appear in the proceedings of the XXXVIIth Rencontres de
    Moriond:
    
    QCD AND HIGH ENERGY HADRONIC INTERACTIONS
\end{center}
\vspace*{1cm}
\title{On the role of the strange quark and its mass
in the chiral phase transition}

\author{J.R. Pel\'aez}

\address{Dip. di Fisica. Universita' degli Studi, Firenze, 
and INFN, Sezione di Firenze, Italy\\
Departamento de
F\'{\i}sica Te\'orica II,  Universidad Complutense. 28040 Madrid,
Spain.}

\maketitle\abstracts{
The evolution of the chiral condensate
with the temperature is studied 
 using SU(3) Chiral Perturbation Theory and the virial
expansion. We observe a decrease of the melting temperature 
of the non-strange condensate
compared with the SU(2) case. Due to the larger mass of the strange quark
we also find an slower temperature evolution
of the strange condensate compared with the  non-strange condensate.}

There is a growing interest 
in the  QCD phase diagram, in the 
transition from a hadron gas to a quark gluon
plasma and in the existence of a chiral phase transition. 
In this note we review a 
rather simple approach \cite{yo} to describe the evolution
of the chiral condensate, namely, 
the virial expansion \cite{virial} of a dilute gas made 
of interacting pions, kaons and etas. Up to second order,
for most thermodynamic properties it is enough to know
the low energy  scattering phase shifts of the particles,
which could be taken from experiment.
However, in order to study the
quark condensates, i.e. derivatives of the 
pressure with respect to the quark masses, a theoretical
description of the mass dependence 
of the scattering amplitudes is needed. 
For that purpose  we turn to
Chiral Perturbation Theory (ChPT) \cite{chpt}, which
provides a remarkable description of 
low energy hadronic interactions. 
Within  ChPT the pions, kaons
and the eta are identified as the Goldstone Bosons of
the QCD spontaneous
chiral symmetry breaking (pseudo Goldstone bosons, 
due to the small quark masses). 
The ChPT Lagrangian is the most general 
derivative and mass expansion respecting the
symmetry constraints built out of $\pi's$, K's and the $\eta$. 
At one loop any calculation can be renormalized
in terms of a finite set of parameters, $L_k$,
 $H_k$.

When SU(2)
ChPT is applied to a gas where only the pions have interactions \cite{Gerber}, 
the phase transition critical temperature can be estimated to be 
$T_c\simeq190\,$MeV.  It has also been shown that the perturbative
calculation is analogous
to the second order virial expansion if the interacting
part of the second virial coefficient is obtained from
 the one loop $\pi\pi$ scattering lengths. 

Recently\cite{yo} we have extended this approach to the SU(3) case,
in order to study the effect of the strange quark and its
large mass. Indeed, the chiral phase transition can be rather different 
for the SU(2) and SU(3) cases, and a larger condensate 
thermal suppression is expected 
with an increasing number of light flavors \cite{flavors}, since, intuitively, the
existence of more states favors entropy and disorder
at the expense of the ordered (condensed) phase.
It is estimated that this so-called ``paramagnetic'' effect 
lowers the $T_c$ down to $150\;$ MeV \cite{lattice} (in the chiral limit).
Within our approach we will check this results from the
hadronic phase with the physical values of the masses.
In addition, within SU(3) we can  study the 
$\langle\bar{s}s\rangle$ condensate. In particular, let us recall that
quark masses play  the same role as magnetic fields
in ferromagnets. As we need 
a higher temperature
to disorder a ferromagnet when there is a magnetic field
along the direction of the magnetization, so it 
has been found that the SU(2) chiral condensate
melts at a higher temperature when taking masses into account\cite{Gerber}.
Since the strange quark mass $m_s >m_u, m_d$, 
we expect this ``ferromagnetic'' effect to raise the
critical $\langle\bar{s}s\rangle$ temperature compared to that of
 $\langle\bar{q}q\rangle$.

For a gas made of
three species: $i=\pi,K,\eta$, 
the pressure relativistic virial expansion 
\cite{Dashen,virial} is :
\begin{equation}
\beta P=\sum_i B_{i}(T)\xi_i + 
\sum_i\left( B_{ii}\xi_i^2 + \frac{1}{2}\sum_{j\neq i}
B_{ij}\xi_i\xi_j
\right)...,
\label{virialexp}
\end{equation}
where $\beta=1/T$ and $\xi_i=\exp({-\beta m_i})$. Expanding
up to the second order in $\xi_i$ means that we only consider
binary interactions. For a free boson gas, the first and second
coefficients are
\begin{eqnarray}
B_i^{(0)}= \frac{g_i}{2\pi^2}\int_0^{\infty} dp\, 
p^2 e^{ -\beta (E(p)- m_i)},
B_{ii}^{(0)}=\frac{g_i}{4\pi^2}\int_0^{\infty} dp\, 
p^2 e^{ -2\beta (E(p)-m_i)},
B_{ij}^{(0))}=0 \;\hbox{for}\; i\neq j ,
\end{eqnarray} 
where the degeneracy is $g_i=3,4,1$ for $\pi, K, \eta$, respectively.
The interactions enter through \cite{Dashen,Gerber}:
\begin{eqnarray}
B_{ij}^{(int)}=\frac{\xi_i^{-1} \,\xi_j^{-1}}{2\pi^3}\int_{m_i+m_j}^{\infty} dE\, E^2 K_1(E/T) 
\sum_{I,J,S} (2I+1)(2J+1)\delta^{ij}_{I,J,S}(E),
\label{delta}
\end{eqnarray}
where $K_1$ is the first modified Bessel function
and the $\delta^{ij}_{I,J,S}$ are the $ij\rightarrow ij$
elastic scattering phase shifts of a state
$ij$ with quantum numbers $I,J,S$ ($J$ being the total angular momentum
and $S$ the strangeness).
Since we are interested in $T<250\,$MeV, it is enough
to consider $ij=\pi\pi,\pi K$ and $\pi \eta$ in the second virial
coefficients.

Quark masses enter the free energy density $z$, as $m_q \bar{q}q$.
Since in $z$ only the pressure depends on the temperature, 
i.e., $P=\epsilon_0- z$
($\epsilon_0$ being the  vacuum energy density) we have:
\begin{equation}
\langle\bar{q}_\alpha q_\alpha\rangle=\frac{\partial z}{\partial m_{q_\alpha}}
=\langle 0 \vert\bar{q}_\alpha q_\alpha\vert 0 \rangle-  
\frac{\partial P}{\partial m_{q_\alpha}}.
\label{condmq}
\end{equation}
where $q_\alpha=u, d,s$.
Note that at $T=0$ we recover 
$\langle\bar{q}_\alpha q_\alpha\rangle=\langle 0 \vert\bar{q}_\alpha q_\alpha\vert 0 \rangle=\partial \epsilon_0/\partial m_{q_\alpha}$. 
Let us emphasize that for
the chiral condensate {\it we  need
the dependence} of $\delta(E)$ {\it on the quark masses} as well as a value
for the vacuum expectation value. For that information we turn to ChPT,
that deals with $\pi's$, K's and $\eta$, and thus we 
translate eq.(\ref{condmq}), in terms of meson masses:
\begin{equation}
\langle\bar{q}_\alpha q_\alpha\rangle
=\langle 0 \vert\bar{q}_\alpha q_\alpha\vert 0 \rangle 
\left(1+\sum_i \frac{c^{\bar{q_\alpha}q_\alpha}_i}{2 m_i F^2} 
\frac{\partial P}{\partial m_i}\right),\quad \hbox{with}\quad
c^{\bar{q_\alpha}q_\alpha}_i=- F^2\frac{\partial m_i^2}
{\partial m_{q_\alpha}}\langle 0 
\vert\bar{q}_\alpha q_\alpha\vert 0 \rangle ^{-1}.
\end{equation}
where, as before, $i=\pi,K, \eta$. The $c$ formulae
to one loop can be easily obtained from
those of  $\langle 0\vert\bar{q} q\vert 0 \rangle$,
$\langle 0\vert\bar{s} s\vert 0 \rangle $ and the 
one loop dependence of the meson masses 
on the quark masses, given a long time ago in \cite{chpt}.
The only relevant comment is that the $c$ depend on
the chiral parameters $L_k$ , for $k=4...8$, and $H_2$.
There are several $L_k$ and $H_2$  determinations
from meson data and  it makes little difference for our conclusions 
which one to use. For further details see ref.\cite{yo}.

Finally, we want $\partial P/\partial m_i$, and for
that we need the theoretical description of the 
meson-meson scattering phase shifts. The meson-meson amplitudes
to one-loop in SU(3) ChPT have been given in \cite{chpt,o4,angelyo}.
These expressions have to be projected in partial waves
of definite isospin $I$ and angular momentum $J$. The complex phases
of these partial waves are the $\delta_{I,J}(E)$ in eq.(\ref{delta}).

The results using the one-loop
ChPT amplitudes, both in SU(3) and SU(2),
 can be seen in Fig.1 \cite{yo} .
The virial coefficients and 
$\partial P/\partial m_i$ have been calculated numerically.
Note that in the isospin limit the $u$ and $d$ condensates
are equal, and we have used the notation
$\langle 0\vert\bar{q} q\vert 0 \rangle \equiv\langle 0\vert\bar{u} u+
\bar{d} d\vert 0 \rangle$.
For the figures, we have represented the 
chiral condensate over its vacuum expectation value,
$\langle \bar{q}_\alpha q_\alpha\rangle /\langle 0 \vert\bar{q}_
\alpha q_\alpha\vert 0 \rangle $,
so that all of them are normalized to 1 at $T=0$. For further
details of the calculation (parameters, etc... see \cite{yo} ).
The melting temperature estimates are:
$T_m^{\langle \bar{q} q\rangle, SU(2)}= 231^{+30}_{- 10}\,\hbox{MeV}$,
$T_m^{\langle \bar{q} q\rangle, SU(3)}= 211^{+19}_{- 7}\,\hbox{MeV}$ and 
$T_m^{\langle \bar{s} s\rangle}= 291^{+37}_{-35}$. 
Note that all these temperatures are strongly correlated, i.e.
the larger in SU(2), the larger in SU(3), 
in particular, we find: 
$T_m^{\langle \bar{q} q\rangle, SU(2)}-T_m^{\langle \bar{q} q\rangle, SU(3)}=21^{+14}_{- 7}\,\hbox{MeV}$. 
This in good agreement with the lattice results \cite{lattice}
in the chiral limit.
About 5 of these MeV are due to free kaons or  etas\cite{Gerber}. 
However the rest of the effect
is due to the different SU(3) and SU(2) meson mass one-loop dependence
on the quark masses. In Figure 1 we have separated the size of 
the different contributions. Furthermore, we can see that 
$\langle \bar{s} s\rangle$ melts slower
than  $\langle \bar{q} q\rangle$. In particular, 
$T_m^{\langle \bar{s}s\rangle}- T_m^{\langle \bar{q}q\rangle}= 
80^{+25}_{-40}\,\hbox{MeV}$. We have also estimated the effect of adding
other, more massive, free hadrons and their contribution decreases 
$T_m^{\langle \bar{q}q\rangle}$ by 7-12 MeV, and $T_m^{\langle \bar{s}s\rangle}$
by 15-23 MeV.

Of course, with the second order virial approach we cannot 
generate a singularity in $T$ like that 
associated to a phase transition. Indeed, our curve does enter 
into the negative region, whereas, in reality, and due to the small
explicit symmetry breaking caused by the quark masses, the 
condensate should only vanish in the $T\rightarrow\infty$ limit. Still, we 
give the complete melting temperature as a 
reference to ease the comparison between different 
curves and previous works. 

\begin{figure}[h]
  \begin{center}
\includegraphics[scale=.61]{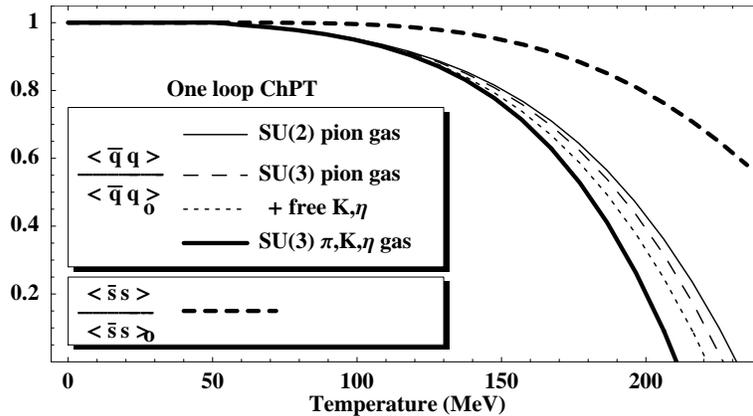}
\vspace{-.5cm}
\caption{\rm Temperature evolution of the quark condensates
  from the one loop ChPT amplitudes (extrapolated to zero for reference).}
\end{center}
\end{figure}
Let us recall that ChPT is an effective approach that 
is only valid at low temperatures
so that beyond 100 or 150 MeV the curves in Fig.1 should be 
considered as a plausible  extrapolation. However, note that both the
 ferromagnetic and paramagnetic effects are already visible at 
low temperatures. It is possible, however, to extend ChPT
up to $E\simeq1.2\,$GeV by means of unitarization methods 
\cite{IAM,angelyo}.
These techniques resum the ChPT series respecting unitarity but
also the low energy expansion, {\em including the mass terms}.
In particular, it has been shown that the coupled channel
Inverse Amplitude Method (IAM) provides an accurate description
of the complete meson-meson interactions
 below 1.2 GeV, generating dynamically
six resonances (and their associated poles): 
$\sigma$, $f_0$, $\rho$, $a_0$, $\kappa$, $K^*$, from
the one-loop ChPT expansion. This approach can be 
extended systematically to higher orders 
(see the two last references in \cite{IAM}). To estimate the 
effect of higher energies, we will use the 
IAM fit\cite{angelyo} to meson-meson data, which was also
 shown to give very accurate scattering 
lengths (only dependent on the masses) with a set of fitted $L_k$ compatible
with previous determinations. 

In Figure II we show the results using the IAM phase shifts.
The continuous line corresponds to the central values, 
and the shaded areas cover the one standard deviation uncertainty
due only to the statistical errors in the parameters from a MINUIT IAM fit.
These areas have been obtained from a Montecarlo gaussian sampling of
$L_k$ and $H_2$ within their errors.
To obtain a conservative estimate of the error 
we have also sampled a conservative range provided in \cite{angelyo}
which estimates some systematic uncertainties but that does not
respect any correlation between chiral parameters.
In such case, the uncertainty is covered by the area
between the dotted lines.
The SU(3) IAM estimate of the melting  temperatures are
$T_c^{\langle \bar{q} q\rangle, SU(3)}
= 204^{+3}_{-1}\,\,(^{\,13}_{5})\, \hbox{MeV}$
and
$T_c^{\langle \bar{s} s\rangle}
= 304^{+39}_{-25}\,\,(^{\,120}_{65})\, \hbox{MeV}$, 
where the errors in parenthesis are the conservative range. 
Again, all temperatures are strongly correlated and 
we find $T_c^{\langle \bar{q} q\rangle, SU(2)}-T_c^{\langle \bar{q} q\rangle, SU(3)}=31.50^{+1.29}_{-0.03}\,\,(^{\,9}_{8})\, \hbox{MeV}$.
The effect of the higher energies seems to be an slight decrease
of all temperatures by about 5 MeV.
Once more we find a slower evolution of the strange condensate
due to the strange quark larger mass, in particular, we find
$\Delta T_c=T_c^{\langle \bar{s}s\rangle}-
T_c^{\langle \bar{q}q_\rangle}= 
100^{+36}_{-29}\,\,(^{\,120}_{80})\,\hbox{MeV}$.
\begin{figure}[h]
  \begin{center}
\includegraphics[scale=.47]{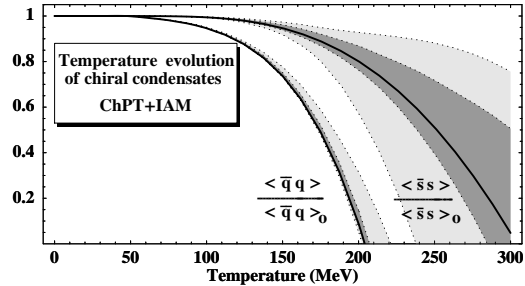}
\end{center}
\caption{\label{fig:epsart} \rm Quark condensate temperature evolution
  from IAM ChPT amplitudes. The shaded areas cover the uncertainties
  in the chiral parameters. ( Curves extrapolated to zero only for
  reference).}
\end{figure}

To summarize, we have studied the SU(2) and SU(3) 
temperature evolution of the chiral condensates {\it in the
hadronic phase}. The thermodynamics of the hadron gas has been 
obtained from the virial expansion and Chiral Perturbation Theory.
Our results clearly show a significant 
decrease, about 20 MeV,  of the non-strange condensate critical temperature,
from the SU(2) to the SU(3) case. 
In addition, they suggest an
slower melting of the strange condensate, 
shifted by 70-80 MeV with respect to the non-strange condensate,
 due to the different quark masses.

\vspace{.1cm}
Work partially supported by Spanish CICYT projects
FPA2000 0956 - BFM2000 1326, and CICYT-INFN collaboration grant.

\bibliography{apssamp}

\end{document}